\documentclass[usenatbib]{mnras}
\usepackage{natbib,amssymb,amsmath,times,graphicx,url,aas_macros}
\usepackage{setspace}      % To manually set line spacing
\usepackage{epstopdf}       % to convert eps images to pdf images
\usepackage{verbatim}       % to allow multiline comments
\usepackage{bm}		  % allow shorthand to bold text
\usepackage{fleqn}		  % flush left equations
\usepackage{Wurster_macros}	 % flush left equations

\newcommand\model[4]{\text{#4}${M}_{#1}\beta_{#2}B_\text{#3}$}
\newcommand\imodel[3]{i${M}_{#1}\beta_{#2}B_\text{#3}$}
\newcommand\nmodel[3]{n${M}_{#1}\beta_{#2}B_\text{#3}$}
\newcommand\hmodel[2]{h${M}_{#1}\beta_{#2}$}

\newcommand\betar[1]{$\beta_\text{r,0} = #1$}
\newcommand\betarz{$\beta_\text{r,0}$}
\newcommand\Machz{$\mathcal{M}_\text{0}$}

\defcitealias{WursterLewis2020d}{Paper I}

\title[Non-ideal MHD vs Turbulence II: Stellar Cores]{Non-ideal magnetohydrodynamics vs turbulence II: \\ Which is the dominant process in stellar core formation?}

\author[Wurster \& Lewis]{James Wurster$^{1,2}$\thanks{jhw5@st-andrews.ac.uk} and Benjamin T. Lewis$^{3}$  \\
$^{1}$Scottish Universities Physics Alliance (SUPA), School of Physics and Astronomy, University of St. Andrews, North Haugh, St Andrews, Fife KY16 9SS, UK \\
$^{2}$School of Physics and Astronomy, University of Exeter, Stocker Rd, Exeter EX4 4QL, UK \\
$^{3}$School of Physics and Astronomy, Rochester Institute of Technology, Rochester N.Y. 14623, U.S.A.
}
\date{Submitted: Revised: Accepted: }
\pagerange{\pageref{firstpage}--\pageref{lastpage}} \pubyear{2020}

\begin{document}
\label{firstpage}
\bibliographystyle{mnras}
\maketitle

%Note: 250word count limit
%203 words
\begin{abstract}
Non-ideal magnetohydrodynamics (MHD) is the dominant process.   We investigate the effect of magnetic fields (ideal and non-ideal) and turbulence (sub- and transsonic) on the formation of protostars by following the gravitational collapse of 1~M$_\odot$ gas clouds through the first hydrostatic core to stellar densities.  The clouds are imposed with both rotational and turbulent velocities, and are threaded with a magnetic field that is parallel/anti-parallel or perpendicular to the rotation axis; we investigate two rotation rates and four Mach numbers.  The initial radius and mass of the stellar core are only weakly dependent on the initial parameters.  In the models that include ideal MHD, the magnetic field strength implanted in the protostar at birth is much higher than observed, independent of the initial level of turbulence; only non-ideal MHD can reduce this strength to near or below the observed levels.  This suggests that not only is ideal MHD an incomplete picture of star formation, but that the magnetic fields in low mass stars are implanted later in life by a dynamo process.  Non-ideal MHD suppresses magnetically launched stellar core outflows, but turbulence permits thermally launched outflows to form a few years after stellar core formation. 
\end{abstract}


%----------------------------------------------------------------------------------------------------------------
\section{Introduction}
\label{intro}

During the low-mass star formation process, gas first gravitationally collapses to the first hydrostatic core;  at the end of this phase, the core undergoes a rapid second collapse until stellar densities are reached and the protostar is born \citep{Larson1969}.  Although the resulting stellar core is a few stellar radii, the cloud from whence it is born is several thousand of au across, and these clouds are embedded in even larger molecular clouds.  Thus, many scales are important in the star formation process.

The molecular cloud is host to many processes, including turbulent velocities \citepeg{Larson1981,HeyerBrunt2004} and magnetic fields \citepeg{HeilesCrutcher2005,Crutcher2012}.  On the cloud-scale, turbulence is supersonic \citepeg{Larson1981}, but has likely decayed to subsonic speeds on the core scale \citepeg{Myers1983,JijinaMyersAdams1999,BerginTafalla2007}.  The cores themselves have been observed to have a uniform rotation \citepeg{Goodman+1993,Caselli+2002}, whose rotation likely originated from the turbulent motion on the larger scales \citepeg{GoodwinWhitworthWardthompson2004a,GoodwinWhitworthWardthompson2004b,Bate2012,Bate2018,WursterBatePrice2019}.  Nonetheless, the actual gas motion in cores is mostly likely a superposition of random (i.e. turbulent) and coherent (i.e. rotational) motions.  

The molecular clouds are only weakly ionised \citepeg{MestelSpitzer1956,NakanoUmebayashi1986,UmebayashiNakano1990}, resulting in interactions between neutral and charged ions.  This is described by non-ideal magnetohydrodynamics \citep[MHD; e.g.][]{WardleNg1999,Wardle2007}, where the important terms for star formation are Ohmic resistivity, ambipolar diffusion and the Hall effect.  In addition to influencing the evolution of a collapsing cloud, magnetic fields also complicate the turbulent gas motion \citep[see the review by][]{HennebelleInutsuka2019}.

There have been many investigations regarding how turbulence and magnetic fields affect the first hydrostatic core phase and the formation and evolution of discs and outflows \citepeg{MatsumotoHanawa2011,Seifried+2012,Seifried+2013,Joos+2013,Myers+2013,Tsukamoto+2015oa,Tsukamoto+2015hall,WursterPriceBate2016,MatsumotoMachidaInutsuka2017,Tomida+2017,WursterBatePrice2018hd,LewisBate2018,GrayMckeeKlein2018,Vaytet+2018};  these studies require the use of sink particles or evolve the disc for a very short period of time.  When using laminar initial conditions and ideal MHD, the magnetic braking catastrophe \citepeg{AllenLiShu2003} prevents the formation of discs.  Using non-ideal MHD or turbulence recovers the discs \emph{under certain initial conditions}, thus in some circumstances, non-ideal MHD or turbulence can prevent (or at least weaken) the magnetic braking catastrophe.  

Given the effect turbulence and non-ideal MHD have on the formation and subsequent evolution of protostellar discs, what effect will they have on the formation of the initial protostar itself?  

Numerically modelling the gravitational collapse from cloud scales all the way to formation of the protostar is challenging since this process spans at least 17 orders of magnitude in density and similar ranges in spatial and temporal scales.  Given the high densities and small dynamical timescales of the resulting stellar core, simulations are only able to model the evolution of the stellar core for a short time after its formation.  Typical end times range from \sm1-2~wks \citep{Machida+2006,MachidaInutsukaMatsumoto2006} to \sm1~mo \citep{Vaytet+2018} to a year \citep{Tomida+2013} to a few years (\citealp{BateTriccoPrice2014,WursterBatePrice2018sd,WursterBatePrice2018hd,WursterBatePrice2018ff}; this paper).  The notable exceptions are \citet{Machida2014} and \citet{MachidaBasu2019} who evolved their stellar cores for \sm270 and 2000~yr, respectively.  Most of these simulations include magnetic fields (including none, some or all of Ohmic resistivity, ambipolar diffusion and the Hall effect), however, turbulence is excluded.  

The simulations that modelled ideal MHD or included Ohmic resistivity formed second core outflows that were launched simultaneously with the birth of the protostar.  These outflows are fast with speeds of $\sim\mathcal{O}(10)$ - $\mathcal{O}(100)$~\kms{} depending on the initial conditions, included physical processes and the integration time after protostar formation; all of these outflows were magnetically launched.   The simulations that included ambipolar diffusion \citep{Vaytet+2018} or Ohmic resistivity, ambipolar diffusion and the Hall effect \citep{WursterBatePrice2018sd,WursterBatePrice2018hd} found that second core outflows were suppressed. 

Under the assumption of ideal MHD and laminar gas flows, magnetic field strengths in excess of 10~kG are embedded in the protostar at its birth \citepeg{Machida+2006,Tomida+2013,BateTriccoPrice2014}, which is higher than the observed kG field strengths around young, low-mass stars \citepeg{JohnskrullValentiKoresko1999,JohnskrullValentiSaar2004,YangJohnskrull2011}.  Including non-ideal MHD in the laminar gas flows decreased the initial magnetic field strength to below observed levels \citep{WursterBatePrice2018ff}.  This led to a resolution of the debate of the origin of magnetic fields in low-mass stars; the two possible origins of the kG-strength surface magnetic fields are that the fields are a `fossil' field that is implanted during the star formation process \citep{Tayler1987,Moss2003,ToutWickramasingheFerrario2004,YangJohnskrull2011}, or that the initial magnetic field is quickly diffused and replaced later by a dynamo-generated field \citep{ChabrierKuker2006}.  The results of \citet{WursterBatePrice2018ff} concluded that the latter theory was correct and further suggested that ideal MHD is an incomplete description of star formation.

While turbulence may solve the magnetic braking catastrophe under certain conditions, can it also prevent unrealistic magnetic field strengths from being implanted in protostars at birth?  

In this study, we investigate the competing effects of non-ideal MHD and sub/transsonic turbulence on the formation of isolated, low-mass protostars using a 3D self-gravitating, smoothed particle, radiative, non-ideal magnetohydrodynamics code.  We follow the collapse through 17 orders of magnitude in density so that our protostar is resolved.    In a companion paper, \citet{WursterLewis2020d} (hereafter \citetalias{WursterLewis2020d}), we follow the collapse through 10 orders of magnitude in density and include sink particles to investigate the effects of turbulence and non-ideal MHD on the formation of a protostellar disc.  In Sections \ref{sec:methods} and \ref{sec:ic} of this paper, we summarise our methods and initial conditions, respectively.  We present our results in \secref{sec:results:core} and conclude in \secref{sec:conclusion}.

%----------------------------------------------------------------------------------------------------------------
\section{Methods}
\label{sec:methods}

Our methods are nearly identical to that which we present in \citetalias{WursterLewis2020d}; the only difference is we exclude sink particles in the current study.  We solve the self-gravitating, radiation non-ideal magnetohydrodynamics equations using the three-dimensional smoothed particle hydrodynamics (SPH) code \textsc{sphNG}.  The code originated from \citet{Benz1990}, but now includes variable smoothing lengths \citep{PriceMonaghan2007}, individual time-stepping \citep{BateBonnellPrice1995}, flux-limited diffusion radiative transfer \citep{WhitehouseBateMonaghan2005,WhitehouseBate2006}, magnetic fields \citep[for a review, see][]{Price2012}, and non-ideal MHD \citep{WursterPriceAyliffe2014,WursterPriceBate2016}.  For stability of the magnetic field, we use the source-term subtraction approach \citep{BorveOmangTrulsen2001}, constrained hyperbolic/parabolic divergence cleaning \citep{TriccoPrice2012,TriccoPriceBate2016}, and the artificial resistivity as described in \citet{Phantom2018}.  For more details, see \citet{WursterBatePrice2018sd}.

We calculate the non-ideal MHD coefficients using version 1.2.3 of the \textsc{Nicil} library \citep{Wurster2016} using the default values detailed in that paper.  At low temperatures ($T \lesssim 600$K), collisions and cosmic rays are the ionisation sources, while at high temperatures the gas is primarily thermally ionised.  The non-ideal effects become unimportant at high temperatures, however, for completeness, we always include these calculations. All non-ideal MHD calculations include Ohmic resistivity, ambipolar diffusion, and the Hall effect.  

%----------------------------------------------------------------------------------------------------------------
\section{Initial conditions}
\label{sec:ic} 

Our initial conditions are identical to \citetalias{WursterLewis2020d}.  In summary, a sphere of mass $M =$ 1~\Msun{}, radius $R=4\times10^{16}$~cm and initial sound speed $c_\text{s} = 2.2\times10^4$~\cms{} is embedded in a box of edge length $L = 4R$; the box and sphere are in pressure equilibrium and have a density contrast of 30:1.  The sphere is given a super-position of solid-body rotation about the $z$-axis (i.e. $\bm{\Omega}_0 = \Omega_0\hat{\bm{z}}$) and a turbulent velocity field; the turbulent field is calculated similarly to \citet{OstrikerStoneGammie2001} and \citet{BateBonnellBromm2003} and described in more detail in \citetalias{WursterLewis2020d} and \citet{LewisBate2018}.  The entire domain is threaded with a magnetic field of strength $B_0=1.63\times10^{-4}$~G = 163~$\mu$G; in the sphere, this is equivalent to five times the critical mass-to-flux ratio (i.e. \mueq{5}) and a ratio of magnetic-to-gravitational energy of $\beta_\text{mag,0}=0.071$.  In all models, the initial ratio of thermal-to-gravitational energy is $\alpha_0 = 0.36$.  The equations for rotational-to-gravitational, turbulent-to-gravitational, magnetic-to-gravitational and thermal-to-gravitational potential energy are given and briefly discussed \citetalias{WursterLewis2020d}.

We include $10^6$ equal mass SPH particles in the sphere and an additional $5\times10^5$ particles in the surrounding medium.  

%----------------------------------------
\subsection{Parameter space}
We investigate the same parameter space both here and in \citetalias{WursterLewis2020d}:
\begin{enumerate}
\item \emph{Magnetic processes}:  We investigate pure hydrodynamics, ideal MHD and non-ideal MHD.  All the non-ideal MHD models include Ohmic resistivity, ambipolar diffusion and the Hall effect.
\item \emph{Magnetic field direction}: For ideal MHD, we investigate the two directions of $\bm{B}_0 = -B_0\hat{\bm{x}} \equiv B_\text{-x}$ and $-B_0\hat{\bm{z}} \equiv B_\text{-z}$.  For non-ideal MHD, we investigate $\bm{B}_0 = -B_0\hat{\bm{x}}$, $-B_0\hat{\bm{z}}$ and $+B_0\hat{\bm{z}} \equiv B_\text{+z}$ since the Hall effect is dependent on the sign of $\bm{\Omega} \cdot \bm{B}$ \citepeg{BraidingWardle2012acc}.
\item \emph{Turbulent Mach number}: We investigate sub- and transsonic values of $\mathcal{M}_0 = $ 0, 0.1, 0.3 and 1.0, corresponding to ratios of turbulent-to-gravitational energy of $\beta_\text{turb,0} = 0$, $0.0012$, $0.011$ and $0.12$, respectively.  In low-mass cores, supersonic values cause a large part of the cloud to unbind, preventing a useful investigation \citep{LewisBate2018}. 
\item \emph{Rotation}: We investigate rotation rates of $\Omega_0 = 1.77\times 10^{-13}$ and $3.54\times 10^{-13}$~s$^{-1}$, corresponding to ratios of rational-to-gravitational energy of $\beta_\text{rot,0} = 0.005$ and $0.02$, respectively.  The former matches the value used in our previous studies and the latter matches the peak of the observed distribution of rotation rates \citep{Goodman+1993}.  These rotations are referred to slow and fiducial, respectively.
\end{enumerate}

Our magnetised models are named \model{\emph{b}}{\emph{c}}{\emph{d}}{\emph{a}}, where $a$ = i (n) for ideal (non-ideal) MHD, $b$ is the Mach number, $c$ is the initial ratio of rotational-to-gravitational energy, and $d$ represents the orientation of the initial magnetic field (i.e. $\pm z$ or $-x$); our hydrodynamic models are named \hmodel{\emph{b}}{\emph{c}}.  An asterisk, *, in place of a variable indicates every model with the remaining defined components.
%----------------------------------------------------------------------------------------------------------------
\section{Results}
\label{sec:results:core}

Following from our studies that investigated the effect of non-ideal MHD effects on the stellar core \citep{WursterBatePrice2018sd,WursterBatePrice2018ff}, we now investigate the effect of including turbulence.  As in our previous studies, we define the birth of the protostar to be at \rhoxeq{-4} and all the gas with \rhoge{-4} to be in the stellar core.  Due to the high densities and consequently very short timesteps, we evolve the ideal models to at least \sm0.75~yr after core formation\footnote{\imodel{0.0}{0.02}{-x} was only evolved to 0.5~yr.} and the remaining models to \sm4~yr after core formation.  

The maximum densities at the end of the simulations are 0.05-0.15~\gpercc, requiring the shortest timestep to represent \sm7~s of real time.    The limiting timestep is the Courant-Friedrichs-Lewy-like condition (see eqn.~1a and associated discussion in \citetalias{WursterLewis2020d}) due to the high densities and negligible non-ideal MHD effects in the stellar core.  Naturally, without replacing the protostar with a sink particle \citepalias[e.g.][]{WursterLewis2020d} or using lower resolution, we cannot evolve the simulation longer than a few years after the formation of the protostar. 

\begin{figure*}
\centering
\includegraphics[width=\columnwidth]{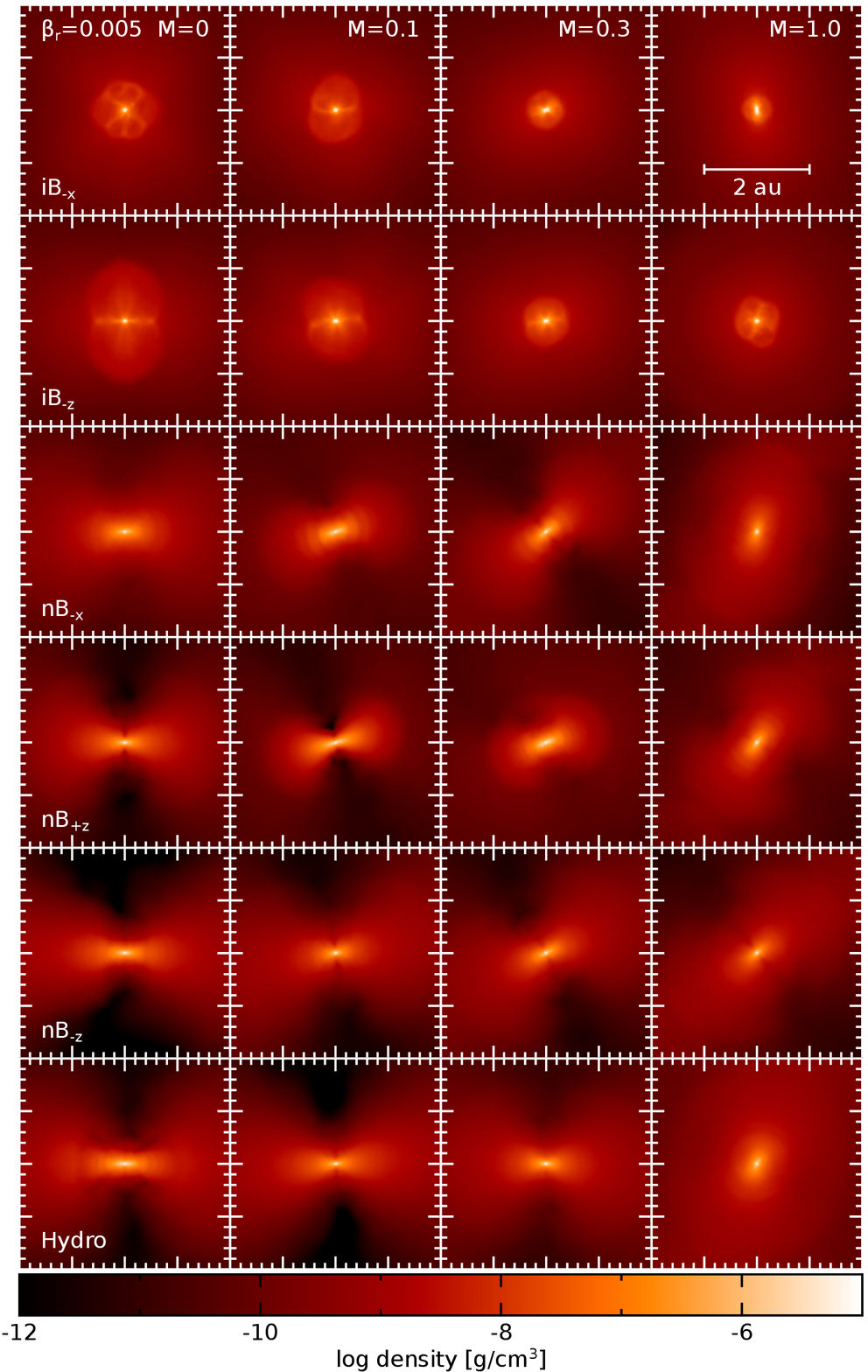}  %Made on Dirac
\includegraphics[width=\columnwidth]{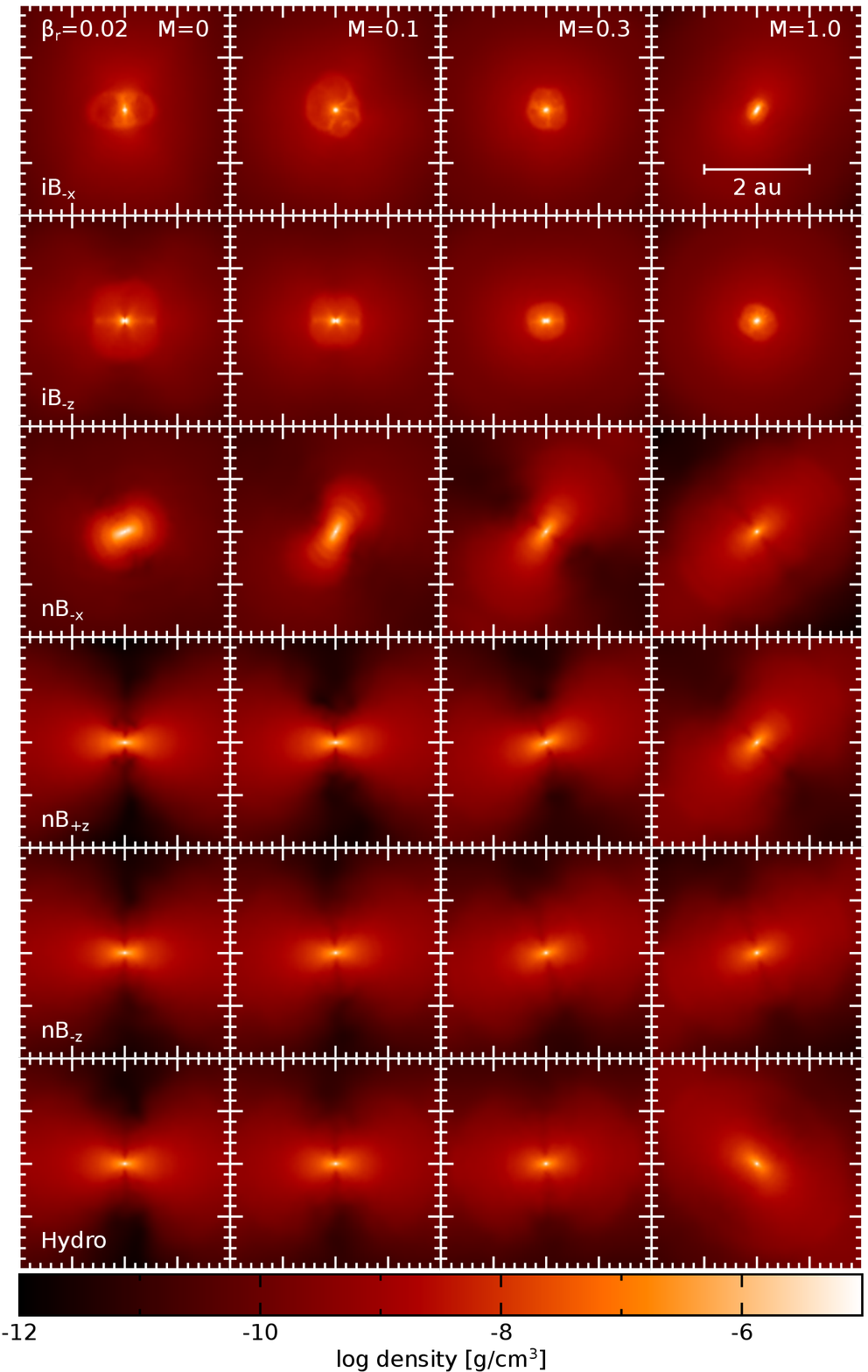}  %Made on Dirac
\caption{Gas density in a cross-section through the centre of the core in the $x$-$z$ plane in the models that use \betar{0.005} (left) and {0.02} (right).  All models have been shifted such that the protostar is at the origin of each frame, but no rotation has been applied.  The non-ideal MHD and hydro frames are at 4~yr after stellar core formation, and the ideal frames are at 0.75~yr after stellar core formation, with the exception of \imodel{0.0}{0.02}{-x} which is at its final time of 0.5~yr.  Increasing the initial Mach number hinders the stellar core outflow in the ideal MHD models, but only misaligns the protostellar disc in the non-ideal and hydro models rather than hindering their formation.}
\label{fig:rho}
\end{figure*}

\figref{fig:rho} shows the gas density in a cross-section through the centre of the protostars at the end of the simulations.  The gas structure in the ideal models is dependent on the initial level of turbulence, such that increasing \Machz{} hinders the stellar core outflow (see \secref{sec:results:core:outflow} below).  A protostellar disc exists in each non-ideal and hydro model, where the turbulence affects its relative angle to the initial rotation axis but not its formation.   The protostellar disc in \nmodel{0.0}{0.02}{-x} is not perpendicular to the initial rotation axis, despite the lack of initial turbulence.  This misalignment has been previously seen in \citet{Tsukamoto+2017} and is a result of the higher initial rotation twisting the magnetic field which in turn causes the misalignment.  By the end of the simulation, its slower rotating counterpart, \nmodel{0.0}{0.005}{-x}, has not twisted the field enough to cause a misaligned disc.  

%----------------------------------------
\subsection{Angular momentum during gravitational collapse}
\label{sec:results:am}

The star formation process is dependent on the angular momentum available.  Previous studies have shown how the angular momentum budget in the first hydrostatic core is related to disc formation \citepeg{Tsukamoto+2015oa,WursterBatePrice2018hd}, suggesting that first cores with more angular momentum are more likely to form discs.  However, \citetalias{WursterLewis2020d} suggested that turbulence did not decrease the angular momentum enough to noticeably affect disc formation.

\begin{figure*}
\centering
\includegraphics[width=\textwidth]{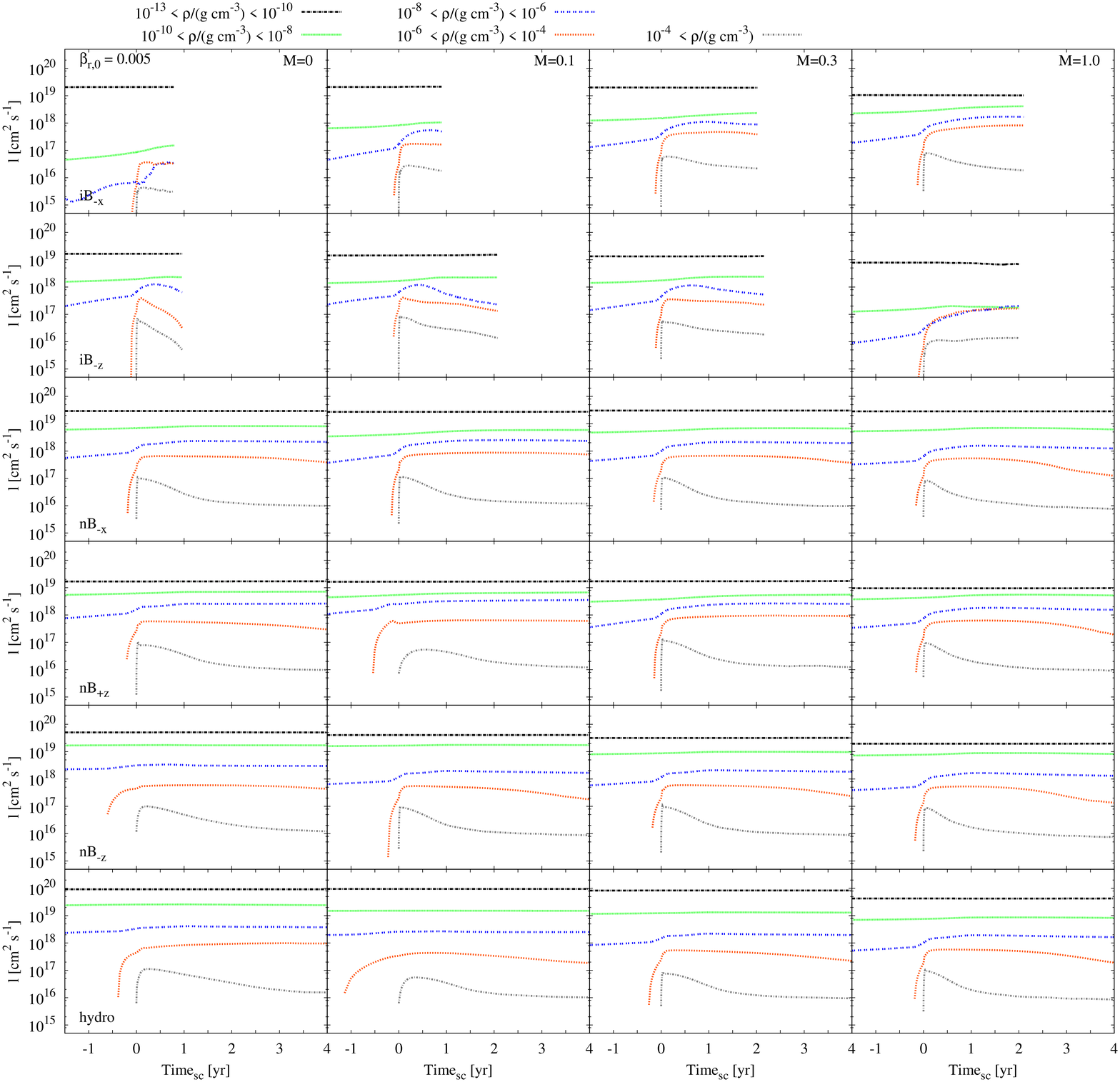}  %Made on Dirac
\caption{Evolution of the specific angular momentum for five density ranges for the models with \betar{0.005}.  Time$_\text{sc} = 0$ represents the formation of the stellar core.  Angular momentum is spread across all density ranges in the non-ideal and hydro models, hindering gravitational collapse, while most of the angular momentum remains at lower gas densities in the ideal models, facilitating collapse.  This demonstrates the efficiency of ideal magnetic fields transporting angular momentum.}
\label{fig:ang:slow}
\end{figure*}

As the first core collapses to form the protostar, the collapsing gas retains some angular momentum, but the amount it contains depends on the initial angular momentum budget and/or the efficiency of the magnetic fields to transport it away from the collapsing gas.  \figsref{fig:ang:slow}{fig:ang:fast} show the evolution of the specific angular momentum in five density ranges in each of our models.  In the ideal models, the angular momentum typically remains in the lower density gas, while in the non-ideal and hydro models, the angular momentum cascades to the higher density gas.  This is a result of ideal magnetic fields efficiently transporting angular momentum away from the centre of the collapsing core.  The more turbulent models tend to also allow the angular momentum to cascade to higher densities, although not with as much efficiency as employing non-ideal MHD.

The angular momentum evolution is similar for both initial rotations, although the models with \betar{0.02} necessarily have more angular momentum at the lower densities.  At higher densities, the specific angular momentum is approximately independent of the initial rotation, suggesting that the initial angular momentum of the stellar core is approximately independent of the initial rotation of the cloud.
\begin{figure*}
\centering
\includegraphics[width=\textwidth]{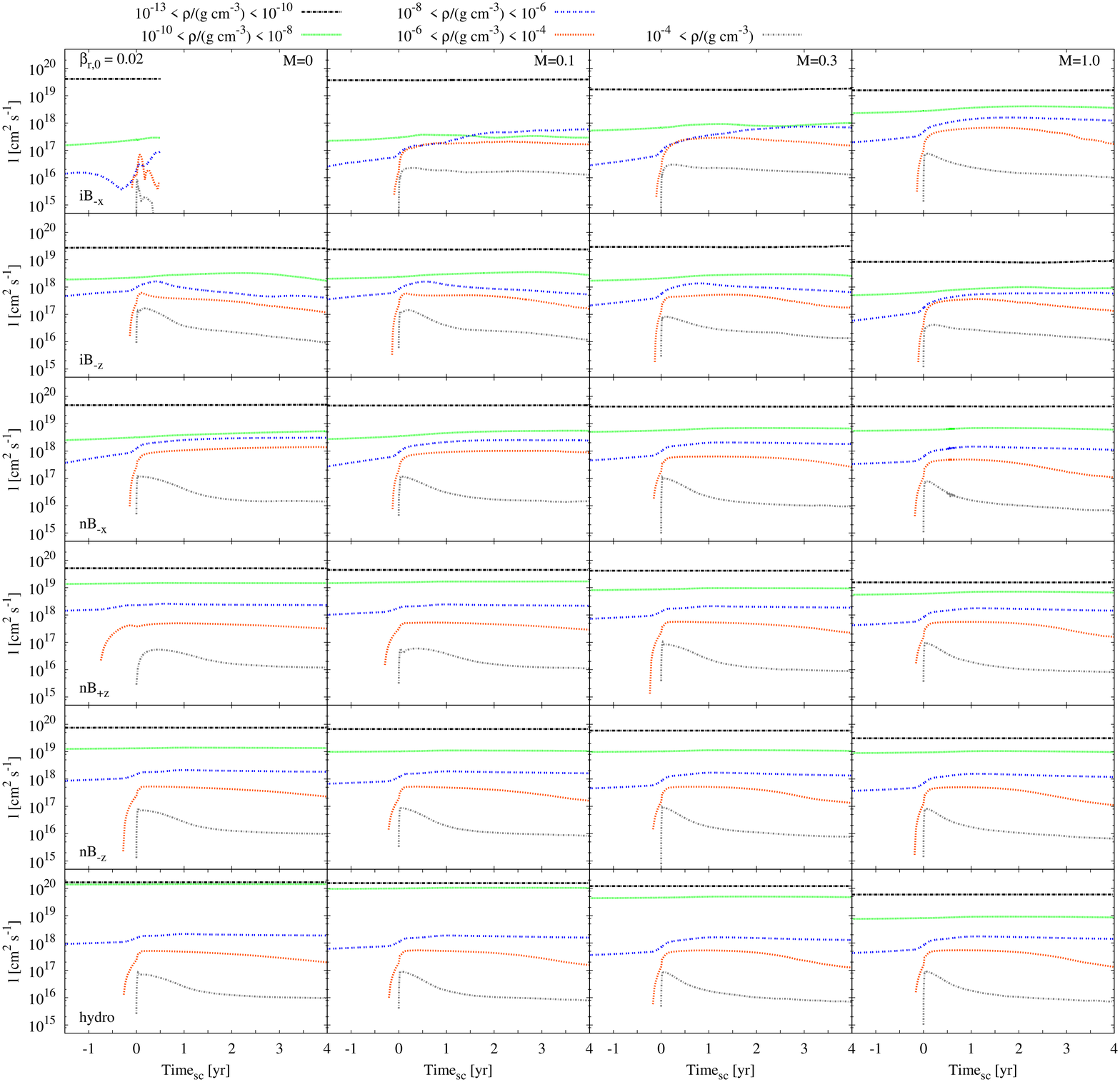}  %Made on Dirac
\caption{The same as \figref{fig:ang:slow} except for \betar{0.02}.  At lower densities, there is more angular momentum than for the lower initial rotation models, while at the higher densities, there is similar quantities of specific angular momentum between the similar models with different initial rotations. This suggests that the initial angular momentum of the cloud does not play an important role in determining the initial properties of the stellar core. }
\label{fig:ang:fast}
\end{figure*} 

%----------------------------------------
\subsection{Stellar core properties}
\label{sec:results:core:prop}

\begin{figure*}
\centering
\includegraphics[width=\textwidth]{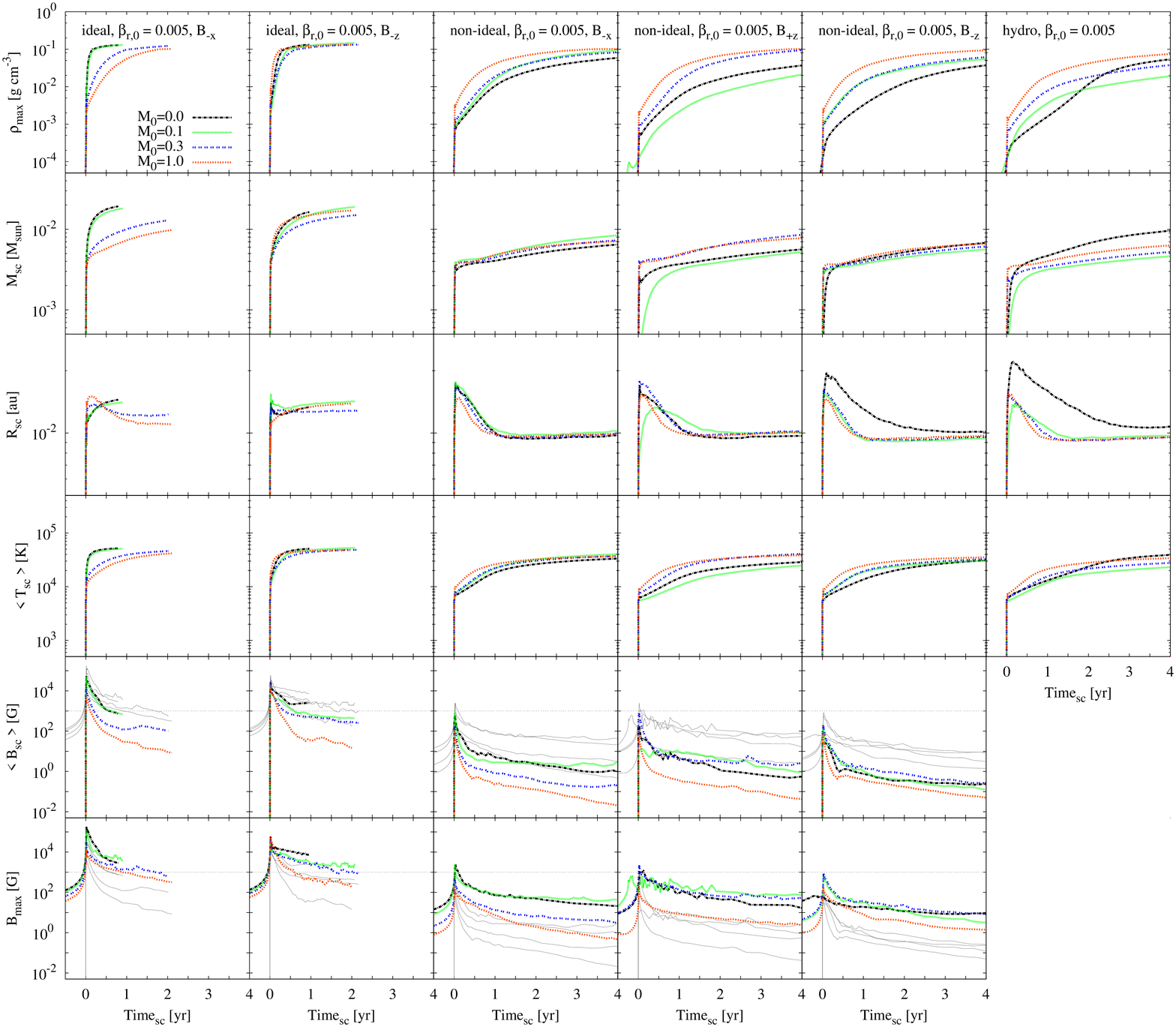}  %Made on Dirac (original version on Linus/Mythos)
\caption{From top to bottom: The maximum density of the stellar core, the mass, radius, average temperature of the stellar core, average magnetic field strength of the stellar core, and the maximum magnetic field strength in the models with \betar{0.005}.   Although turbulence affects the initial properties of the stellar cores, the largest contribution to the differences is the inclusion of non-ideal MHD.  In the magnetic plots, the horizontal line is a reference line at $B = 10^3$~G and the thin grey lines in the $\left< B_\text{sc} \right>$ panels are $B_\text{max}$ for reference and vice versa.  The maximum and core strengths at birth in the ideal MHD models are in excess of the kG-strength magnetic fields found in low-mass stars, suggesting that ideal MHD is a poor approximation when modelling stars, even if turbulence is included.  The core strength in the non-ideal MHD models is $B_\text{core} \lesssim 10^3$~G, indicating that non-ideal MHD is required to realistically model star formation, and further suggesting that magnetic fields in low-mass stars are generated later by a dynamo process. }
\label{fig:sc:slow}
\end{figure*}

\figsref{fig:sc:slow}{fig:sc:fast} show the time evolution of the maximum density, the stellar core mass, radius, average temperature, average magnetic field strength and the maximum magnetic field strength in each model.
\begin{figure*}
\centering
\includegraphics[width=\textwidth]{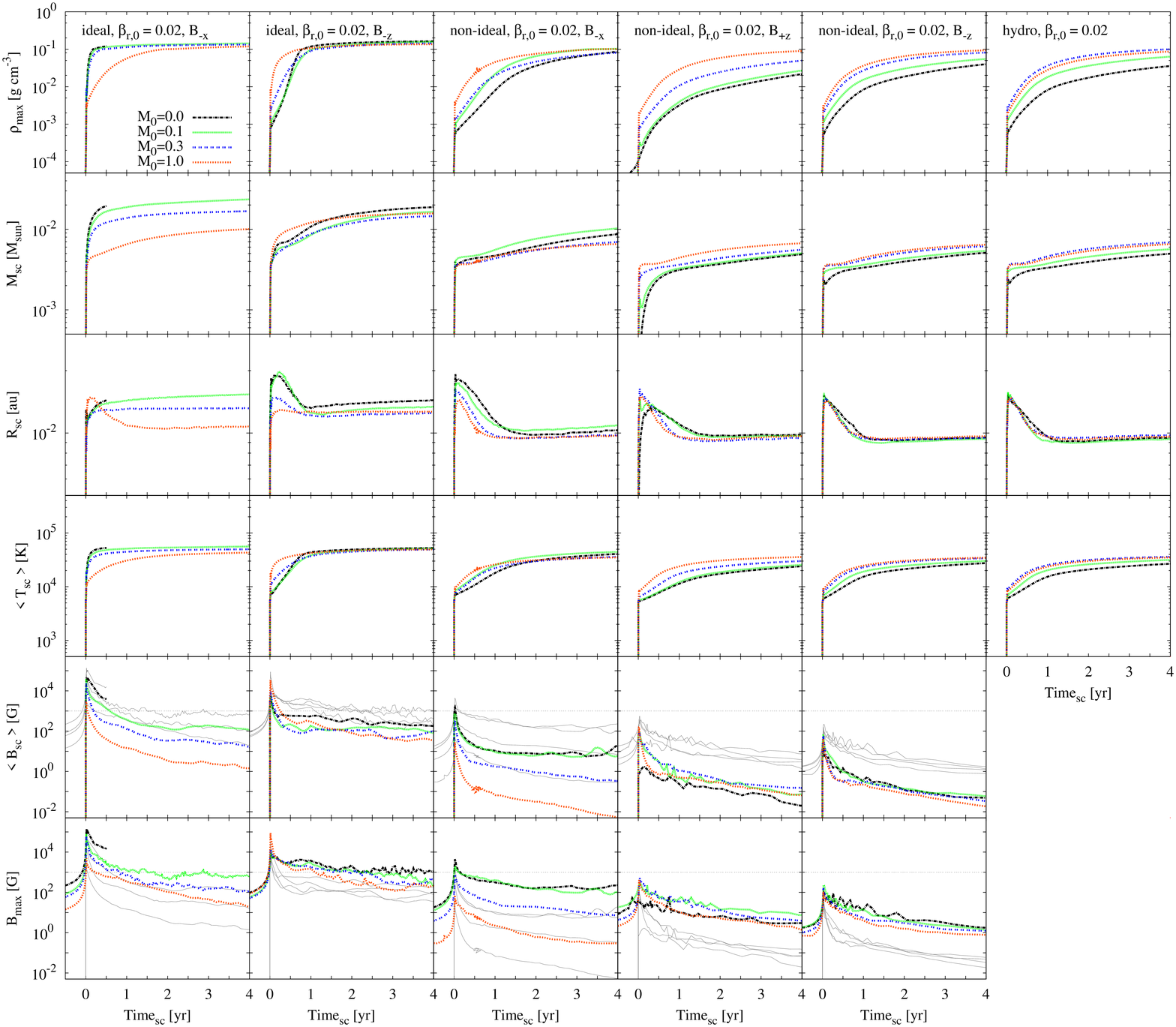}  %Made on Dirac (original version on Linus/Mythos)
\caption{The same as \figref{fig:sc:slow} except for \betar{0.02}.   Increasing the initial rotation rate from \betar{0.005} to \betar{0.02} has a minimal affect on the initial properties of the stellar core.}
\label{fig:sc:fast}
\end{figure*} 

A few ideal MHD models undergo rapid collapse to stellar densities of \rhoxapprox{-1}, although most stall their rapid collapse at \rhoxapprox{-3} and then slowly continue to increase their maximum density.  This decrease in the growth of \rhox{} has previously been seen in the literature when the initial magnetic field strength is decreased \citep{BateTriccoPrice2014} or in non-ideal MHD models when the cosmic ray ionisation rate is increased \citep{WursterBatePrice2018sd}.  This difference in evolution is a result of angular momentum transport, as discussed above.  The models with little specific angular momentum cascading to high densities will rapidly collapse, whereas those whose angular momentum transport is hindered will have slower growth rates.

The growth rate of \rhox{} is mirrored in the growth rates of the stellar core mass (second row of \figsref{fig:sc:slow}{fig:sc:fast}).  Although the ideal models have stellar core masses that are consistently more massive than the non-ideal models, we cannot reach any conclusion given that most of the models are continuing to accreted.  However, amongst the ideal models and independently amongst the non-ideal models, the stellar core masses vary by less than a factor of two at any given time.  This suggests that the initial core mass is not dependent on the initial level of turbulence.

These stellar masses are approximately an order of magnitude lower than those presented in \citetalias{WursterLewis2020d}, however, the two sets of masses are not directly comparable.  The main difference is that these masses represent the `true' mass of the protostar, whereas the masses in \citetalias{WursterLewis2020d} comprise of all the mass within a sphere of 1~au, which is much larger than the `true' protostellar radii.

In most models, the radius (third row) quickly reaches a maximum, and then contracts slightly until it reaches a new equilibrium of $0.01 \lesssim R_\text{sc}/\text{au} \lesssim 0.02$.  During this contraction the specific angular momentum decreases (\figsref{fig:ang:slow}{fig:ang:fast}), suggesting that some angular momentum may be transported outwards, likely to the gas with \rhoxrange{-8}{-6}.  The models that undergo rapid collapse to stellar densities are less likely to undergo the radial contraction, although these cores already have little angular momentum compared to the remaining models.   

The evolution of the stellar core temperature (fourth row of \figsref{fig:sc:slow}{fig:sc:fast}) also reflects the evolution of \rhox{}.  Shortly after formation of the stellar core, the temperature is similar in all models, varying by only a factor of a few.  Thus, the stellar core temperature at birth is approximately independent of initial conditions.

These results suggest that several of the initial properties of the stellar core -- radius, mass, and temperature -- are approximately independent of initial conditions.  Given the similarity of these values to those previously published in the literature \citep[e.g.][these studies collectively span a wide arrange of initial conditions]{Machida+2006,MachidaInutsukaMatsumoto2006,Tomida+2013,Vaytet+2018}, this suggests \emph{all} stellar cores form with similar initial properties.

%-----
\subsection{Magnetic fields}

\begin{figure*}
\centering
\includegraphics[width=\columnwidth]{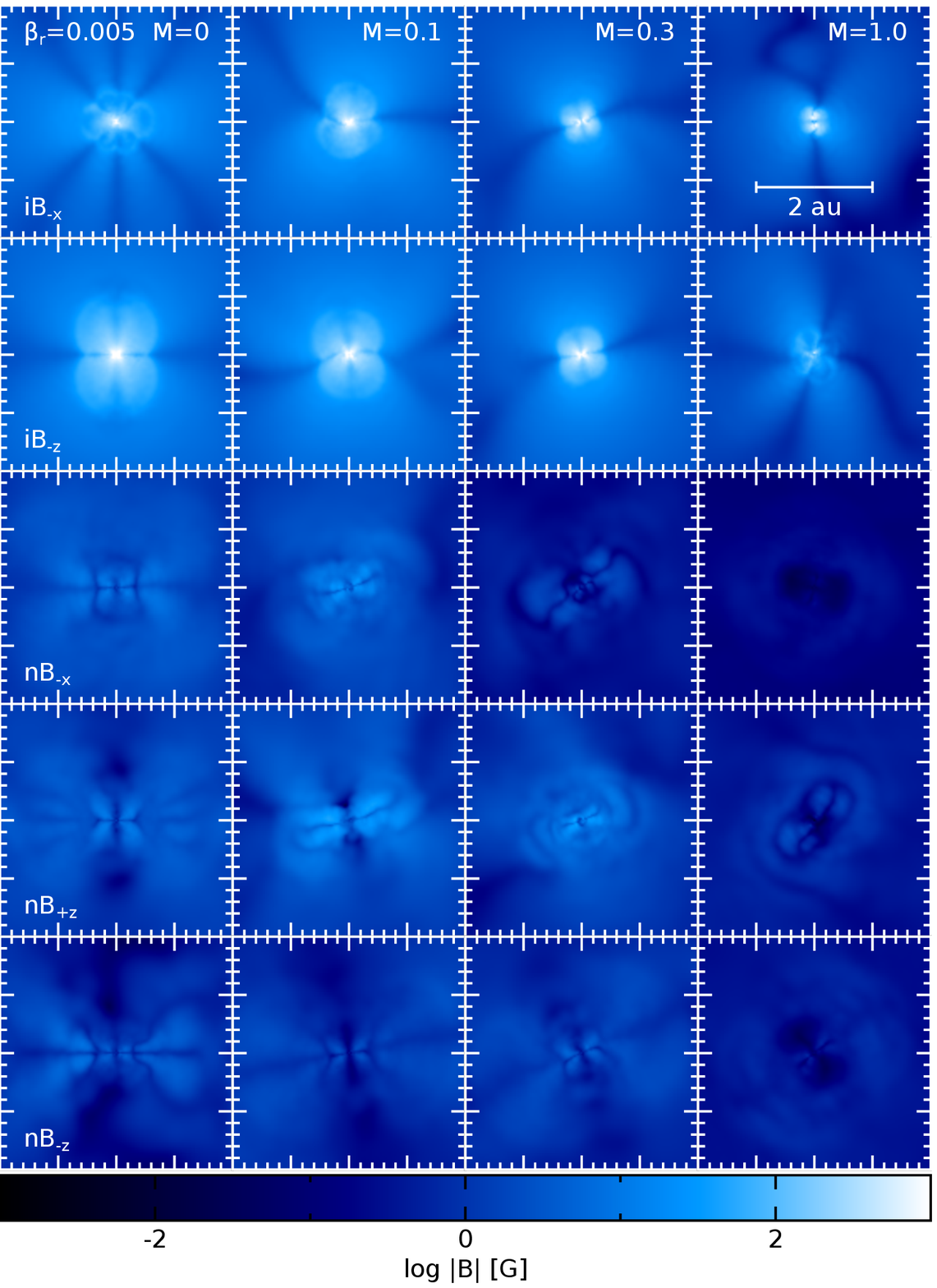}  %Made on Dirac
\includegraphics[width=\columnwidth]{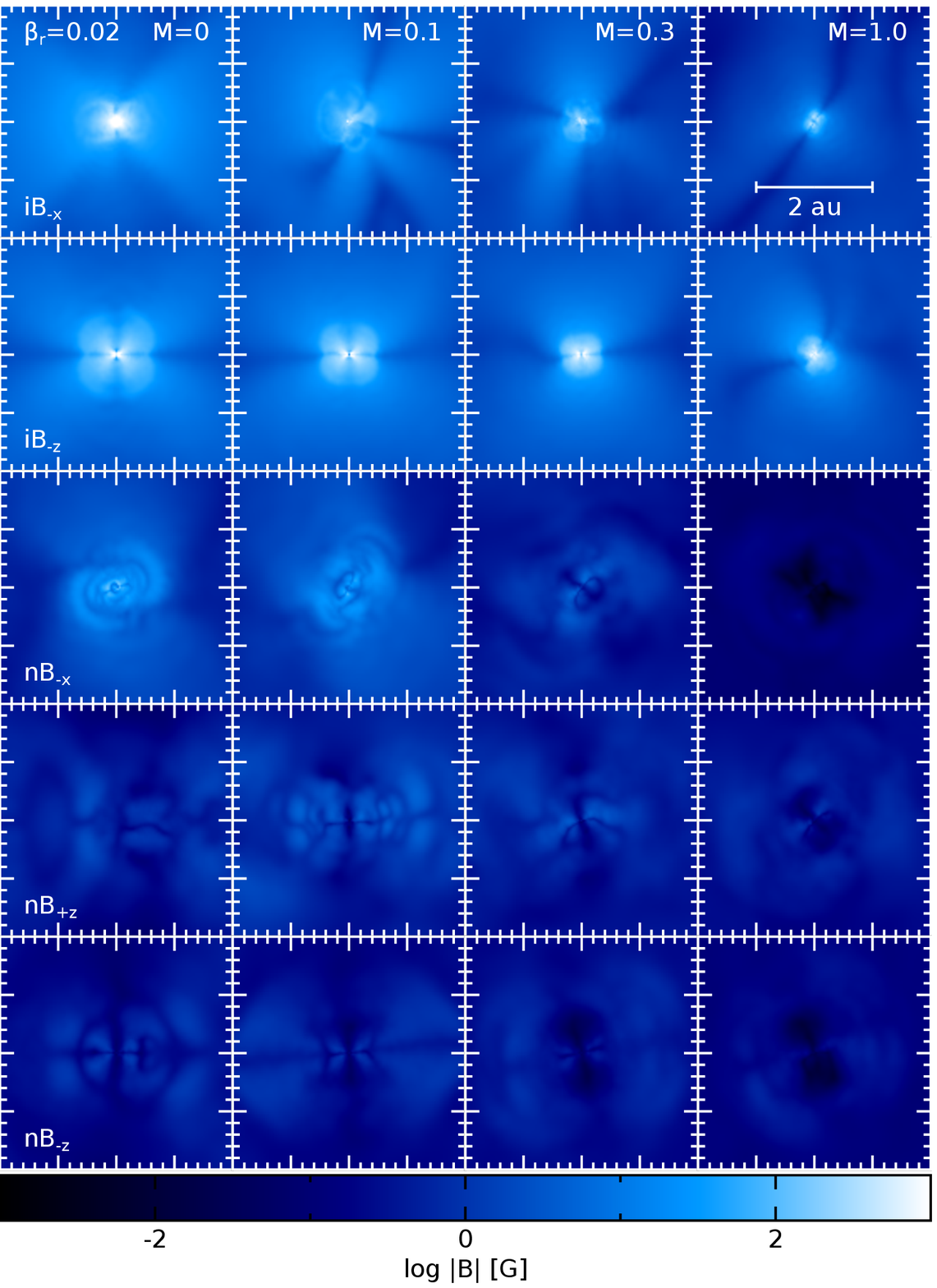}  %Made on Dirac
\caption{Magnetic field strength in a cross-section through the centre of the core in the $x$-$z$ plane the models that use \betar{0.005} (left) and {0.02} (right).  All models have been shifted such that the protostar is at the origin of each panel.  The times are the same as in \figref{fig:rho}.  The magnetic field strength decreases for increasing initial Mach number.  In the ideal models, stronger field strengths are coincident with the stellar core and the outflows, while in non-ideal models, weaker fields are coincident with outflows. }
\label{fig:B}
\end{figure*}

\figref{fig:B} shows the magnetic field strength in a cross-section through the core.  In the ideal models, the prominent features (i.e. the stellar core and outflows) are regions of higher magnetic field strength.  In the non-ideal models, the stellar cores have a weak magnetic field strength, and for increasing in initial Mach number, the field strength around the protostar tends to decrease, with very weak magnetic fields surrounding the protostars in the transsonic models.

\begin{figure}
\centering
\includegraphics[width=\columnwidth]{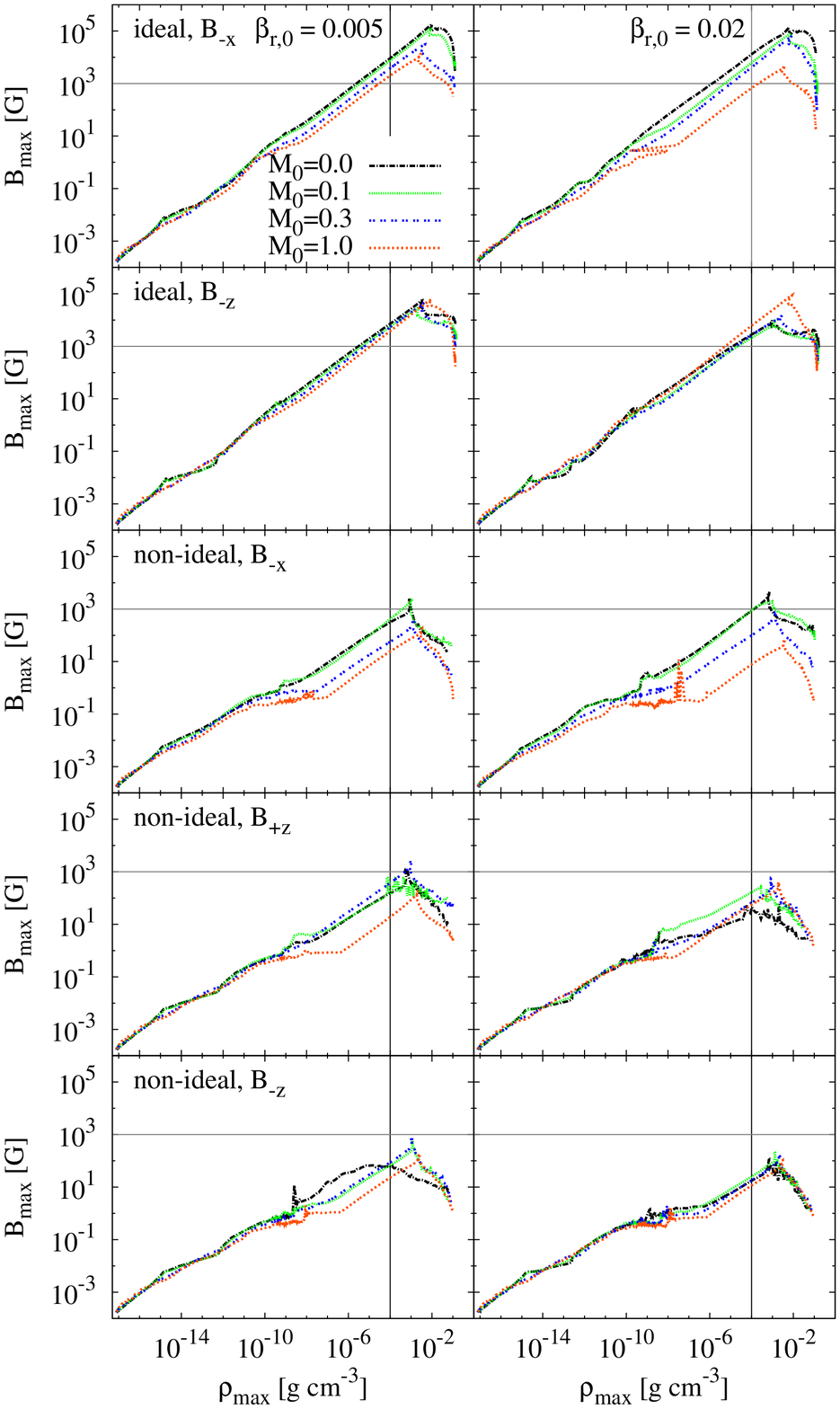}  %Made on Dirac
\caption{The evolution of the maximum magnetic field strength as a function of maximum density for our models.  The vertical line at \rhoxeq{-4} represents the birth of the protostar, and the horizontal line at $B_\text{max} = 10^3$~G represents the observed magnetic field strength in young, low-mass stars.  Including non-ideal MHD has the greatest impact on reducing the maximum magnetic field strength, although in a few models (\imodel{*}{0.02}{-x}, \nmodel{*}{*}{-x}, \nmodel{*}{0.005}{$\pm$ z}), including turbulence also has a noticeable impact on decreasing the magnetic field strength.}
\label{fig:BvsRho}
\end{figure}

The magnetic field of the stellar core at birth necessarily depends on the evolution of the cloud prior to the core's formation \citepeg{Tomida+2013,BateTriccoPrice2014,WursterBatePrice2018sd,WursterBatePrice2018hd,WursterBatePrice2018ff}.   \figref{fig:BvsRho} shows the evolution of the maximum magnetic field strength as a function of maximum density (which is a useful proxy for time, until perhaps the formation of the protostar) for all the models.   In the ideal models, the maximum field strength is located at the centre of the core, whereas in the non-ideal models, the maximum magnetic field strength is located in the gas surrounding the stellar core \citep[][and verified here]{WursterBatePrice2018ff}.

\subsubsection{Maximum magnetic field strength}

The evolution of the maximum magnetic field strength begins to differ during the first hydrostatic core phase, depending on the initial conditions.  Turbulence plays a small role in determining the evolution of $B_\text{max}$ in \imodel{*}{*}{-z}, but does affect the evolution of \imodel{*}{*}{-x}, where $B_\text{max}$ can differ by \sm1.5 orders of magnitude during the second collapse phase ($10^{-8} \lesssim$ \rhox/\gpercc{} $\lesssim 10^{-4}$).  In \imodel{*}{*}{-x}, the amount of angular momentum differs in each density bin amongst the models (\figsref{fig:ang:slow}{fig:ang:fast}), suggesting that each model has a different efficiency at transporting angular momentum.  In the models with the greater distribution of angular momentum amongst the density bins, the field lines are not as easily dragged into the centre of the core, leading to the weaker $B_\text{max}$.

In all ideal models, $B_\text{max} \gtrsim 10^3$~G by the time the stellar core forms.  Although turbulence can decrease the strength of the maximum magnetic field, it cannot decrease it well below the observed $10^3$~G threshold required to determine the origin of magnetic fields in low-mass stars \citepeg{JohnskrullValentiKoresko1999,JohnskrullValentiSaar2004,YangJohnskrull2011}.

The maximum magnetic field strength decreases with time after the formation of the stellar core (see also the bottom row of \figsref{fig:sc:slow}{fig:sc:fast}).  In many of these models, the maximum field remains \sm$10^3$~G, although in a few cases it decreases to 10-100~G.  Due to computational limitations, we cannot evolve these simulations further, thus cannot comment on the extent of this decrease.

In the non-ideal models, the maximum magnetic field strength surpasses $10^3$~G in only a few models.  Thus, non-ideal MHD is more efficient than turbulence at decreasing the magnetic field strength.  For \nmodel{*}{*}{-x}, \nmodel{*}{0.005}{$\pm$ z}, increasing the initial turbulence tends to decrease the maximum field strength during the second collapse by as much as a factor of 100, suggesting that turbulence can amplify the non-ideal effects.   After the stellar core has formed, the field strength decreases, and remains well below the $10^3$~G threshold in all cases.

\subsubsection{Stellar core magnetic field strength}
In the ideal MHD models, the field strength of the stellar core is similar to the maximum field strength since the maximum field strength is at the centre of the core.  When considering the core strength of the ideal models, this value decreases to between \sm10 and 100~G in some of the more turbulent cases shortly after the core is formed.   Therefore, in these models, a dynamo action is required later in life to increase the magnetic field strength to the observed levels, concluding that magnetic fields in low mass stars are not fossil fields.  Although a few turbulent ideal MHD models suggest a conclusion to the dynamo-fossil field debate, we must be cautious since this conclusion depends on the level of turbulence and neglects that the magnetic field strength is $B_\text{max} \sim 10^5$~G at the formation of the stellar core itself.

In the non-ideal models, the maximum magnetic field strength resides outside the stellar core, thus the stellar core field strength is consistently 1-2 orders of magnitude lower than the maximum field strength (compare the bottom two rows of \figsref{fig:sc:slow}{fig:sc:fast}).  This strongly suggests that when non-ideal MHD is included, the magnetic field of the star is low enough to conclude that its origin is from a dynamo action later in life.

Therefore, we can only reach a confident conclusion regarding the origin of magnetic fields in low-mass stars when employing a complete description of all the physical process involved in star formation.  This necessarily means including non-ideal MHD to model all scales.

%----------------------------------------
\subsection{Stellar core outflows}
\label{sec:results:core:outflow}

Stellar core outflows have been launched from laminar, ideal MHD simulations \citepeg{BateTriccoPrice2014,WursterBatePrice2018sd} and laminar simulations with Ohmic resistivity \citepeg{MachidaInutsukaMatsumoto2006,Tomida+2013,MachidaBasu2019}.  When including Ohmic resistivity, ambipolar diffusion and the Hall effect, \citet{WursterBatePrice2018sd,WursterBatePrice2018hd} found that the strength of the stellar core outflow decreased and the outflow ultimately disappeared for models with cosmic ray ionisation rates of $\zeta_\text{cr} = 10^{-16}$ to $10^{-17}$~s$^{-1}$; these models used \betar{0.005}.

\begin{figure*}
\centering
\includegraphics[width=\columnwidth]{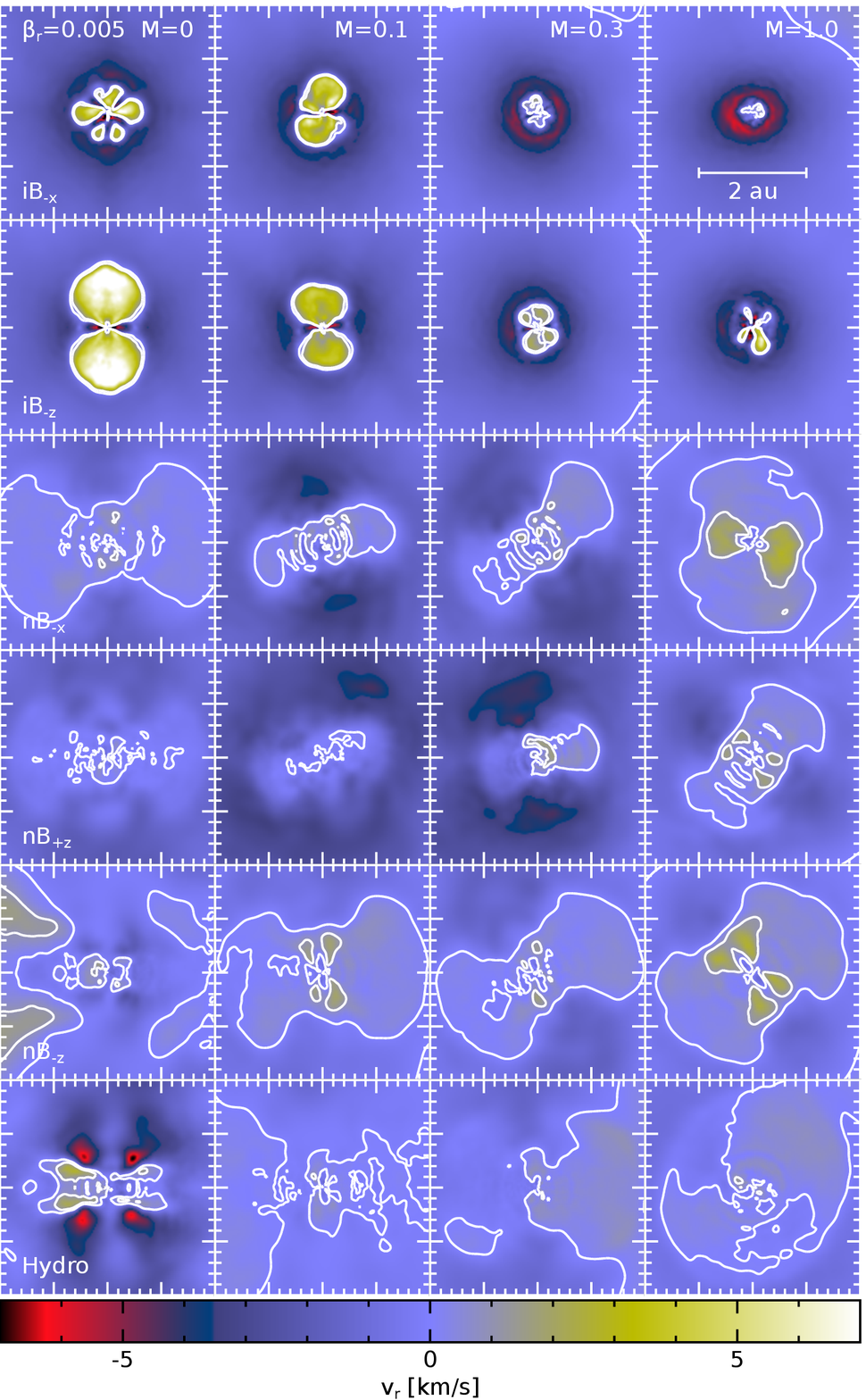}  %Made on Dirac
\includegraphics[width=\columnwidth]{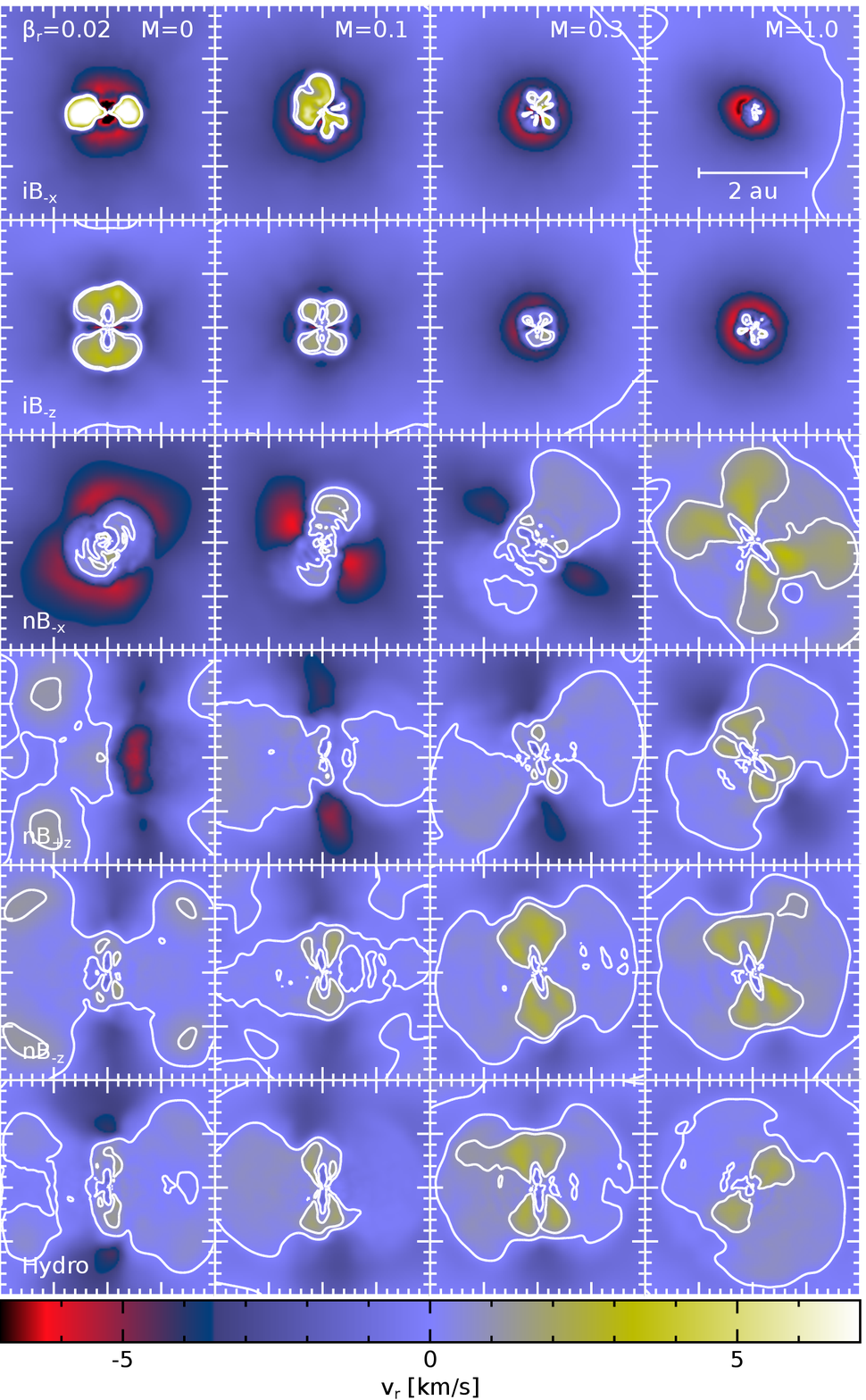}  %Made on Dirac
\caption{Radial velocity in a cross-section through the centre of the core in the $x$-$z$ plane the models that use \betar{0.005} (left) and {0.02} (right).  All models have been shifted such that the protostar is at the origin of each panel.  Contours are at $v_\text{r} = 0, 1$~\kms.  The times are the same as in \figref{fig:rho}.  In the ideal models, increasing the initial turbulence hinders the launching of magnetic outflows simultaneous with stellar core formation.   In the non-ideal models, increasing the initial turbulence promotes the launching of thermal outflows after the stellar core has formed. }
\label{fig:vr}
\end{figure*}

\figref{fig:vr} shows the radial velocity in a cross-section through the stellar cores.  When employing ideal MHD, increasing the initial level of turbulence hinders the formation of outflows, with (e.g.) a fast, magnetically launched outflow in \imodel{0.0}{0.005}{-z}, but nearly no outflows in \imodel{1.0}{*}{-x}.  For non-ideal MHD and hydro, increasing the initial level of turbulence permits outflows to form.

\begin{figure}
\centering
\includegraphics[width=\columnwidth]{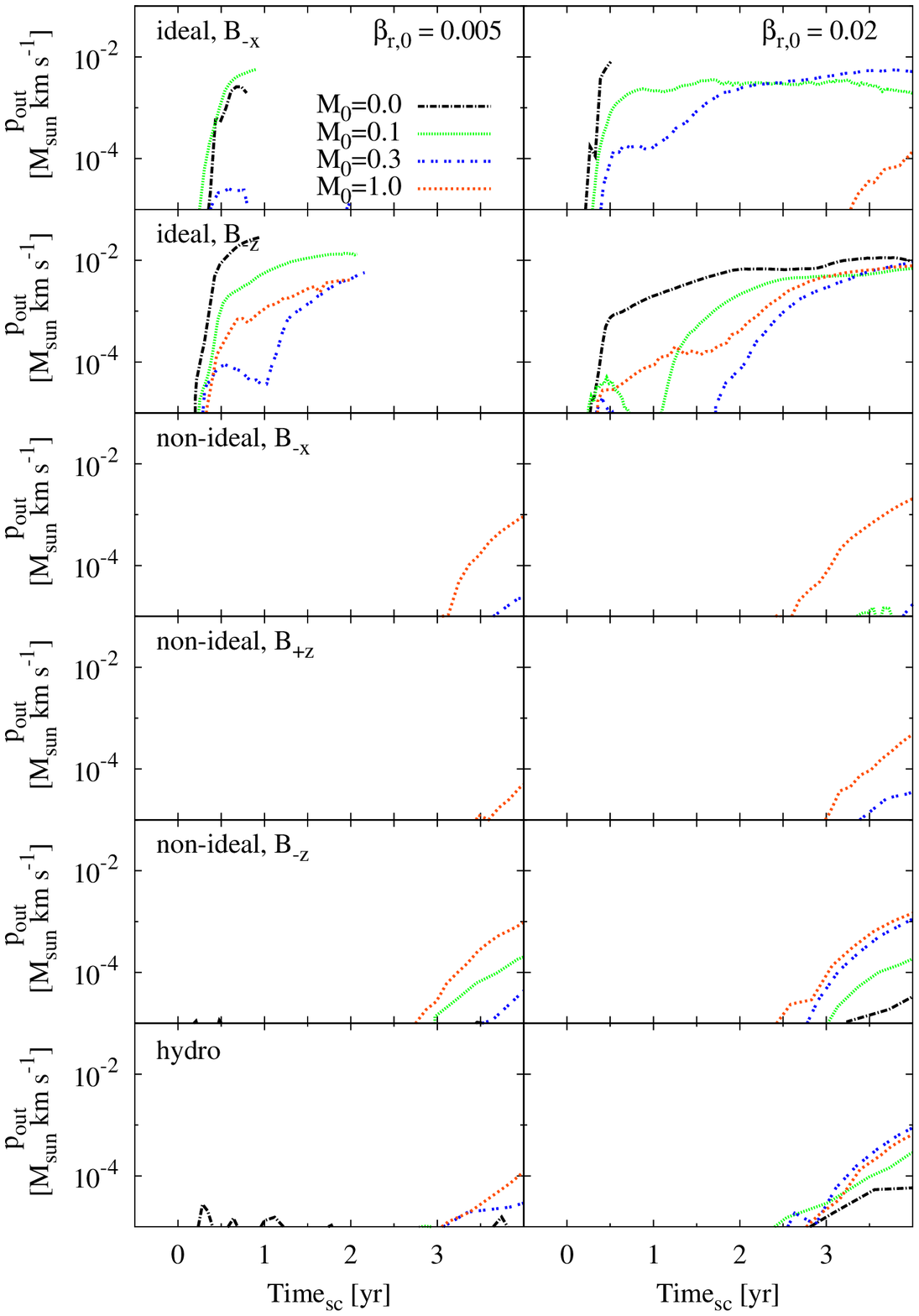}  %Made on Dirac
\caption{The evolution of the amount of momentum, $p=mv_z'$, in the second core outflow.  The outflow consists of the gas within $r < 2$~au of the stellar core, $\rho < 10^{-8}$~\gpercc{} and $v_z'/v > 0.5$, where $v_z'$ is the component of the velocity parallel to the outflow axis; they are either too weak to be resolved or non-existent for $p < 10^{-5}$~\Msun{}~\kms{}.  Outflows are launched early in the ideal models, with some delay resulting from turbulence.  When outflows are launched in the non-ideal and hydro models, they are launched a few years after the formation of the star and tend to contain less momentum than the ideal outflows.}
\label{fig:sc:Pout}
\end{figure}

\figref{fig:sc:Pout} shows the evolution of the amount of momentum in the second core outflows.  We consider the gas to be in the second core outflow if it is within $r < 2$~au of the stellar core, $\rho < 10^{-8}$~\gpercc{} and $v_z'/v > 0.5$, where $v_z'$ is the component of the velocity parallel to the outflow axis.

In most of the ideal models, outflows are launched almost immediately after the formation of the stellar core.  Most of these outflows carry considerable momentum, reaching $10^{-2}$~\Msun{}~\kms{} within a year.  These outflows tend to be magnetically launched and correlate to regions of strong magnetic fields.   As with the first core outflows \citepalias{WursterLewis2020d}, increasing the initial Mach number decreases the collimation of the outflows.  Given the nature of adding randomly seeded turbulence\footnote{Previous studies \citepeg{GoodwinWhitworthWardthompson2004a,Liptai+2017,Geen+2018} have shown how changing the initial seed can affect the results of a simulation.}, there are exceptions, where the outflow is either delayed or suppressed in \imodel{0.3}{0.005}{-x} and \imodel{1.0}{*}{-x}.

In the absence of turbulence, non-ideal MHD suppress stellar core outflows, independent of initial magnetic field orientation and initial rotation.  However, given enough initial rotation or turbulence, then stellar core outflows are recovered.  These outflows are launched \appx3~yr after stellar core formation (which is much later than in the ideal models) and at a slower speed than the in the ideal models (i.e. there is less momentum in the non-ideal and hydro outflows than in the ideal outflows over the first year after they are launched).  For the non-ideal and hydro models, increasing \betarz{} and/or \Machz{} tends to increase the amount of momentum in the outflow, indicating that the initial gas motion affects the stellar core outflows.

Unlike the ideal models, the outflows in the non-ideal models are not correlated to regions of strong magnetic fields.  Specifically, in the regions surrounding the stellar core, the magnetic field typically decreases in strength in the non-ideal models, with the outflows comprised of gas that is more weakly magnetised than the surrounding material.  To analyse the outflow, we compare the radial acceleration due to thermal pressure (i.e. $\left|\text{d}P/\text{d}r\right|/\rho$), to the magnitude of the vertical component of the Lorentz acceleration (i.e. $\left| \bm{J} \times \bm{B}\right|_{z'}/\rho$ where $z'$ is parallel to the outflow axis).   In the ideal models with outflows, the Lorentz acceleration is comparable with the acceleration due to thermal pressure for  $r \lesssim 2$~au.  In the non-ideal models with outflows, the Lorentz acceleration is a few orders of magnitude smaller than the pressure acceleration for $r \lesssim 2$~au.  Thus, these non-ideal outflows are driven by thermal pressure, and are a result of the large amount of thermal energy liberated during first core formation \citepeg{Bate2010,Bate2011,SchonkeTscharnuter2011,BateTriccoPrice2014}.   This is very similar to the hydrodynamic models, where outflows are necessarily driven solely by thermal pressure. This suggests that the environment around a core in a non-ideal MHD model may more closely resemble the hydrodynamic case.

Therefore, turbulence affects the formation of stellar core outflows.  Increasing turbulence decreases the collimation of the magnetically launched outflows when modelling ideal MHD, although it does not significantly impact the total amount of momentum in the outflows.  Increasing turbulence promotes the launching of thermally driven outflows when modelling non-ideal MHD.

%----------------------------------------------------------------------------------------------------------------
\subsection{Resolution}
\label{sec:res}

The numerical formation of protostars is known to be dependent on many physical and numerical processes, including resolution \citepeg{BateTriccoPrice2014,WursterBatePrice2018ff}.  Our resolution of $10^6$ particles was chosen based upon the large suite of simulations presented here and in \citetalias{WursterLewis2020d} and the computational resources available.  Even increasing the number of particles by a factor of 3 would increase the required resources of each model by a factor of \sm10, thus we consciously decided to run the large suites at the current resolution.  Although quantitative results will change with resolution, the consistency amongst the models in our suite means that the relative results will hold, meaning that we can reasonable compare the effect of turbulence versus non-ideal MHD, as presented above.

Given our resolution, there are \sm$10^4$ particles in the stellar core; these particles have smoothing lengths of $h \approx 3\times10^{-4}$ - $10^{-3}$~au.  These smoothing lengths are greater than the minimum cell size of $8\times10^{-5}$~au in \citet{Vaytet+2018}, but smaller than the minimum cell size of $5.6\times10^{-3}$~au in \citet{MachidaBasu2019}, thus our resolution is comparable to that presently in the literature.

\citet{WursterBatePrice2018ff} showed that the maximum and central magnetic field strengths can increase by 2 and 4 orders of magnitude, respectively, in ideal models when increasing from $3\times10^5$ particles to $3\times10^6$.  Our current resolution is $10^6$ particles, thus we expect increasing resolution would increase the field strengths of our ideal models by possibly an order of magnitude.  If so, then the core magnetic field strengths in the ideal models would rise above the observed value of $10^3$~G, reaffirming that ideal MHD -- even with turbulence -- is an incomplete description of star formation.  

When modelling non-ideal MHD (specifically a counterpart to \nmodel{0.0}{0.005}{+z}),  \citet{WursterBatePrice2018ff} found that the central and maximum magnetic field strengths were relatively insensitive to resolution.  Therefore, the magnetic field strengths of the non-ideal models as shown in \figsref{fig:sc:slow}{fig:sc:fast} are reliable values as is the conclusion that non-ideal MHD is required to decrease the magnetic field below observed values and that the magnetic field in stars must be generated by a dynamo later in life.

%----------------------------------------------------------------------------------------------------------------
\section{conclusion}
\label{sec:conclusion}

We have presented a suite of simulations that followed the gravitational collapse of initially rotating 1~\Msun{} gas cores through to stellar densities, with a focus on the effect that turbulence and non-ideal MHD has on the stellar core properties.  In \citetalias{WursterLewis2020d}, we investigated the effect that these processes had on the formation and early evolution of the protostellar disc.  We simulated collapses that were purely hydrodynamical, those that employed ideal MHD and those that employed non-ideal MHD, while varying the Mach number, initial rotation speed, and magnetic field direction.  Once the stellar core was formed, we evolved the system for an additional 0.75-4~yr.  Our key results are as follows:
\begin{enumerate}
\item Several initial properties of the stellar cores -- radius, mass, and temperature -- are approximately independent of all initial conditions.
\item The non-ideal processes are more efficient at decreasing the strength of the magnetic field than turbulence.  Even with turbulence, the magnetic field strength implanted in stars at birth was orders of magnitude higher in  the ideal models compared to the non-ideal MHD models.  The high values of the central and maximum magnetic field strengths in the ideal models indicated that ideal MHD is an incomplete picture of star formation.
\item The protostars that formed in the non-ideal MHD models were implanted with weak $\lesssim$~kG-strength magnetic field at birth, suggesting that the magnetic field in low-mass stars must be generated later by a dynamo process.  This conclusion is independent of the initial level of turbulence.
\item Increasing the initial Mach number decreased the collimation of the stellar core outflows in the ideal MHD models.  These outflows were magnetically launched nearly simultaneously with the formation of the protostar.
\item Increasing the initial Mach number permitted stellar core outflows to be launched in the non-ideal MHD and hydro models.  These outflows were thermally launched \appx3~yr after the formation of the protostar.
\end{enumerate}

Aside from stellar core outflows, the initial level of turbulence has a minimal role in the formation and early evolution of the stellar core, indicating that non-ideal magnetohydrodynamical processes are more important than sub- and transsonic turbulent processes.

%----------------------------------------------------------------------------------------------------------------
\section*{Acknowledgements}

We would like to thank the anonymous referee for useful comments that greatly improved the quality of this manuscript.
JW acknowledges support from the European Research Council under the European Community's Seventh Framework Programme (FP7/2007- 2013 grant agreement no. 339248), and from the University of St Andrews.
BTL acknowledges the support of the National Aeronautics and Space Administration (NASA) through grant NN17AK90G and from the National Science Foundation (NSF) through grants no 1517488 and PHY-1748958.
The authors would like to acknowledge the use of the University of Exeter High-Performance Computing (HPC) facility in carrying out this work.  
Analysis for this work was performed using the DiRAC Data Intensive service at Leicester, operated by the University of Leicester IT Services, which forms part of the STFC DiRAC HPC Facility (www.dirac.ac.uk). The equipment was funded by BEIS capital funding via STFC capital grants ST/K000373/1 and ST/R002363/1 and STFC DiRAC Operations grant ST/R001014/1. DiRAC is part of the National e-Infrastructure.  
The column density figures were made using \textsc{splash} \citep{Price2007}.
%The research data supporting this publication are openly available from the University of Exeter's institutional repository at: https://doi.org/10.24378/exe.XXX.
The research data supporting this publication will be openly available from the University of Exeter's institutional repository.

%----------------------------------------------------------------------------------------------------------------
\bibliography{TurbVsNimhd_cores}
%----------------------------------------------------------------------------------------------------------------

\label{lastpage}
\end{document}